\documentclass[a4paper,11pt]{article}
\usepackage{pos}
\usepackage{graphicx}  
\usepackage{dcolumn}   
\usepackage{bm}        
\usepackage[section]{placeins}
\usepackage{amsmath}
\usepackage{epsfig}
\usepackage{rotating}
\usepackage{xspace}
\usepackage{hyperref}
\usepackage{xcolor}
\usepackage{cleveref}
\newcommand{\bea}{\begin{eqnarray}}
\newcommand{\eea}{\end{eqnarray}}

\title{Complex Langevin boundary terms in full QCD}
\ShortTitle{Complex Langevin boundary terms}

\author*[a]{Michael Westh Hansen}
\author[a]{D\'enes Sexty}

\affiliation[a]{Institute of Physics, NAWI Graz, University of Graz,\\
Universitätsplatz 5, Graz, Austria }

\emailAdd{michael.hansen@uni-graz.at}
\emailAdd{denes.sexty@uni-graz.at}

\abstract{Lattice simulations of non-zero density QCD introduce the so-called sign problem (complex or negative probabilities), which invalidates importance sampling methods. To circumvent this, we use the Complex Langevin Equation (CLE), to measure the boundary terms and then compare these results with the ones gotten from reweighting, confirming the expectations from previous studies. We also investigate boundary terms in simulations using CLE with dynamic stabilization and compare this, to results calculated with reweighting.}

\FullConference{%
The 39th International Symposium on Lattice Field Theory,\\
8th-13th August, 2022,\\
Rheinische Friedrich-Wilhelms-Universität Bonn, Bonn, Germany
}

\begin{document}
\maketitle

\section{Introduction}
The sign problem appears in systems haveing a complex action, invalidating importance sampling simulations. The method discussed in this article is the complex Langevin method \cite{Klauder:1983nn}, which uses a complexified stochastic process, rather than a probability distribution to circumvent the sign problem. For some complex measure $ \rho(x)=\exp(-S(x)) $ depending on the variable $x$ the Complex Langevin equation (CLE) is written as 
\begin{equation}
\partial_\tau x = - \textrm{Re} \left.{ \frac{\partial S(z)}{\partial z} } \right|_{z=x+iy} + \eta_\tau, \qquad \partial_\tau y = - \textrm{Im} \left.{ \frac{\partial S(z)}{\partial z} } \right|_{z=x+iy} 
\label{CLE}
\end{equation}
with the drift term $ K(z) = \partial S(z) / \partial z $ and a Gaussian noise $\eta$ satisfying $ \langle \eta_\tau \eta_{\tau'} \rangle=2 \delta(\tau-\tau') $.
Even though this method was shown to circumvent the sign problem successfully in many models, sometimes converge towards incorrect results is observed.
It has been identified that slow decay of the distribution on the complexified manyfold can lead to this behavior \cite{Aarts:2011ax,Nagata:2016vkn}.
In full QCD the Complex Langevin equation has been show to provide reliable results at small lattice spacings, where it has been used to e.g.
calculate the equation of state in the deconfined state \cite{Sexty:2019vqx,Attanasio:2022mjd}.
The slow-decay problem in this case is fixed by introducing gauge cooling \cite{Seiler:2012wz}. In other cases (at large lattice-spacings or small temperatures) gauge cooling is not sufficient to achieve fast enough decay. The boundary terms were introduced to identify the incorrect results within the CLE simulation\cite{Scherzer:2018hid,Scherzer:2019lrh}. The measurement of the boundary terms are cheap even for a lattice system, since it amounts to measuring a new observable and performing an offline analysis to be detailed below. 
Dynamical stabilization\cite{Attanasio:2018rtq} introduces a gauge invariant force to the drift term with the aim to confine the complexified process close to the original SU(3) manifold, to regularize the system in the cases where gauge cooling is not sufficient. This has shown good results in toy models, and can be applied to QCD as well. 

In this work we study full QCD discretised using the plaquette action and two or four flavors of staggered fermions at nonzero chemical potential, with relatively light quark masses.
We measure boundary terms and test the performance of simulations using Dynamical stabilization by comparing their results to the reweighted results
of HMC simulations at zero $\mu$.

\section{Boundary terms}
The Complex Langevin equation (CLE) creates a real probability density on a complexified manifold $P(x,y,\tau)$, where $\tau$ is known as the Langevin time. (Here we illustrate it for one complexified scalar variable, the generalization to more complicated systems easily follows.)
First let us define an observable in the complexified manifold,
\begin{equation}
    \langle O \rangle_{P(t)} = \int dx dy P(x,y,t) O(x+iy).
\end{equation}
Whereas we denote averages with the complex density $ \rho(x,t) $ as
\begin{equation}
    \langle O \rangle_{\rho(t)} = \int dx O(x) \rho(x,t),
\end{equation}
Here the time dependent distribution is evolving as $\rho(x,t) = \exp(L_c^T) \rho_0(x)$, with $\rho_0(x)$ being some initial distribution, and we 
defined the complex Fokker-Planck operator or Complex Langevin operator, $L_c = (\partial_z + K(z))\partial_z$. 
It can be shown that provided $L_c$ has a spectrum with non-positive real part eigenvalues, $ \langle O \rangle_{\rho(t)} $ converges to 
the correct result $ 1/Z \int dx e^{-S(x)} O(x)$. 
Now having the complex distribution and the correct distribution, we can interpolate between them, allowing for verification of our Complex 
Langevin results.
\begin{equation}
    F_O(t,\tau) = \int dx dy P(x,y,t-\tau)O(x,y,\tau),
\end{equation}
with the time-evolved observable
\begin{equation}
    O(t,\tau) = exp(\tau L_c) O(x,y).
\end{equation}
It is easy to see that $F_O(t, 0) = \langle O \rangle_{P(t)}$, the less obvious $F_O(t, t) = \langle O \rangle_{\rho(t)}$ is also true (provided
assumptions made above hold). This implies that if the $\tau$ derivative of our interpolation function is zero, we have correct results. 
\begin{equation}
    \frac{\partial F(t,\tau)}{\partial \tau}=0
\end{equation}
However this is not always satisfied in practice.
To better understand its behavior we introduce a cut-off, to define:
\begin{equation}
    B(Y,t) = \left. \partial_\tau F_O(Y;t,\tau)\right|_{\tau=0} = \int_{|y|<Y} dx dy \partial_\tau P(x,y,t-\tau) O(x,y)  
\end{equation}
The value of the boundary terms are then to be observed in the $ Y \rightarrow \infty$ limit.
Here we can make use of the Fokker-Planck equation, and change the derivative $\partial_t P(x,y,t) = L_c P(x,y,t)$
\begin{equation}
    B(Y,t) = -\int_{|y|<Y} dx dy \left(\partial_t P(x,y,t) \right) O(x,y) + \int_{|y|<Y} dx dy P(x,y,t) L_c O(x,y)  
\end{equation}
and since the distribution $P(x,y,t)$ settles as $t \to \infty $, the first integral vanishes.
Applying the derivative multiple times to the interpolation function we get higher order boundary terms, which are calculated similarly to the first one. 
\begin{equation}
    B_n(Y,t) = \left. \partial^n_\tau F_O(Y;t,\tau)\right|_{\tau=0} = \int dx dy P(x,y,t) L^n_c O(x,y) \Theta(Y-y).
\end{equation}
Thus the boundary terms are defined in terms of new observables $L_c^n O$, which also include 
a cutoff in the imaginary part of the fields. 
For lattice systems, the definitions are the same, however the cutoff now has to depend on all of the gauge links in order to enclose
a compact submanifold of the complexified manifold. Thus we define a unitarity norm, as a measure of how non-unitary our gauge links are.
\begin{equation}
       n(M) = \textrm{Tr} (M^\dagger M -1 )^2 \textrm{ for } M \in \textrm{SL}(N,\mathbb{C}).
\end{equation}
This can then be used as the cutoff in the boundary term
\begin{equation}
    B_n(Y) = \int P(M) L_c^n O \Theta ( Y-n(M)  )d M.
    \label{bound}
\end{equation}

\section{Results}
First we investigate the plaquettes and Polyakov loops in full QCD, to investigate whether the boundary term can provide a way to confirm correct observables. The boundary term for the plaquette and the Polyakov loop has been calculated using eq. (\ref{bound}), using the left derivative
\begin{equation}
    D_a U = i \lambda_a U   \qquad D_a U^{-1} = - i U^{-1} \lambda_a 
\end{equation}
in the $L_c = \sum_i (D_i + K_i) D_i$ operator, where $\lambda_a$ are the Gell-Mann matrices, and the $i$ index includes space-time coordinates,
Lorentz and color indices. In this section we present results from small ($4^4$) lattices, as our aim was to test the feasibility and validity of the boundary term analysis for full QCD simulations. We use 2 flavors of (rooted) staggered fermions with mass parameter $m=0.02$. The setup of the simulations (regarding discretisation of the 
Langevin eq. for the gauge link variables, gauge cooling, etc) is similar to what 
have been used in \cite{Sexty:2013ica}, except for the calculation of the fermionic drift force, where in this case, instead of a noisy estimator we have used an exact calculation of the fermionic drift term
\begin{equation}
    K^F_i = - N_f \textrm{Tr} (M^{-1} D_i M )
\end{equation}
where $M$ is the staggered Dirac matrix. This allows simulations at low $\beta$ values, as the simulations using noisy estimator tend to be more instable in that region , as also noticed in \cite{Bloch:2017jzi}.
\begin{figure}[h]
\begin{center}
  \includegraphics[width=0.48\columnwidth]{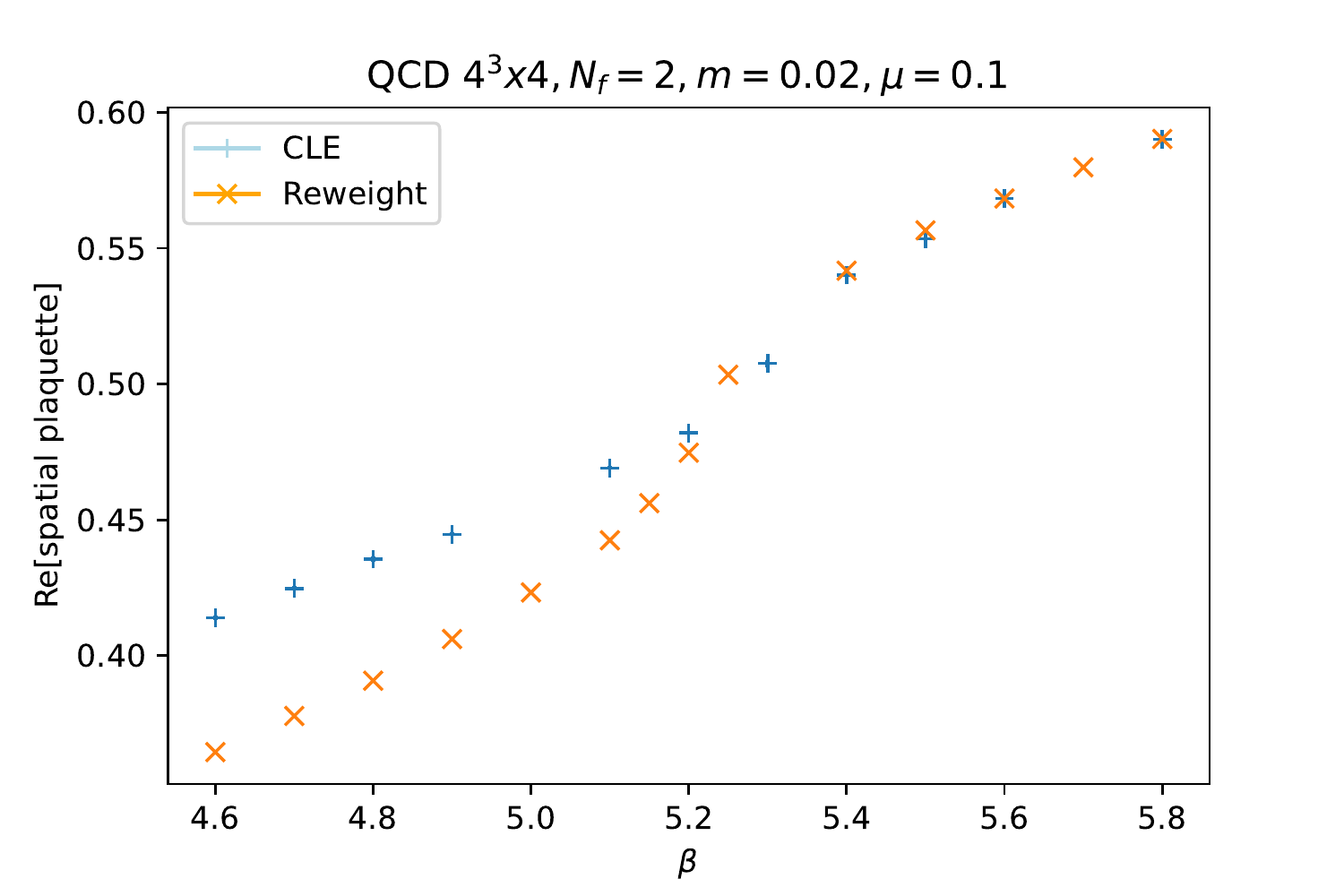}
  \includegraphics[width=0.48\columnwidth]{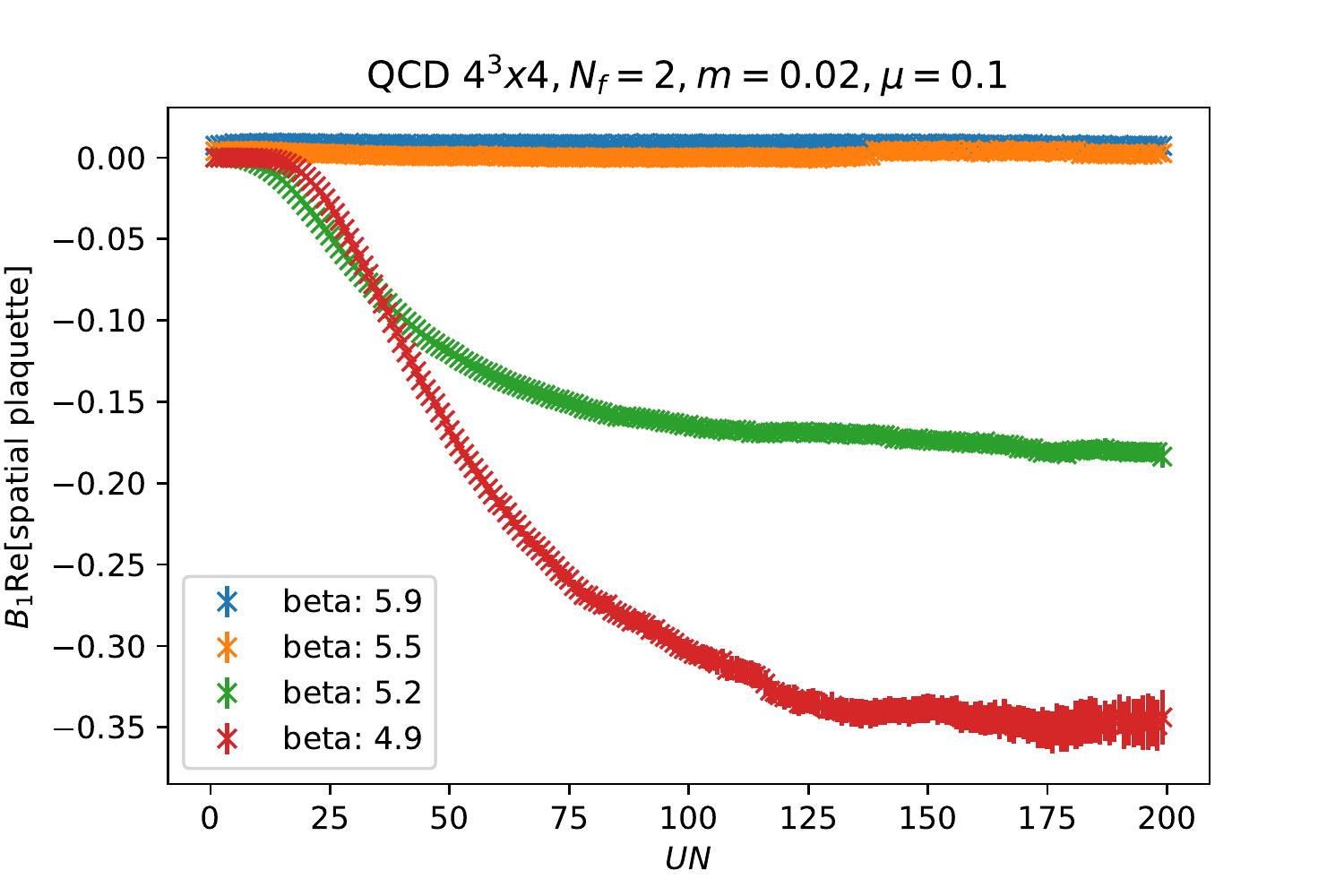}
\caption{On the left panel the Spatial plaquette is plotted, with results calculated from CLE(blue) and reweighted HMC(orange). 
On the right panel the boundary terms are shown for various values of beta as a function of the cutoff. }
\label{sp_mu0.1}
\end{center}
\end{figure}
To compare our CLE data  with correct results, we have used data from a HMC simulations at zero quark chemical potential$(\mu =0)$, reweighted to a non-zero $\mu$. We have used $ \mu = 0.1 $ corresponding to $\mu/T=0.4$, such that reweighting is 
still feasible here. We used O(1000) configurations at each $\beta$ value to perfom the reweighting. 
It can be seen in Fig.~\ref{sp_mu0.1} that the CLE is incorrect for low beta values as observed in earlier studies \cite{Fodor:2015doa}. This can be also correctly observed from the boundary terms, as seen on the right panel of Fig.~\ref{sp_mu0.1}.
\begin{figure}[h]
\begin{center}
  \includegraphics[width=0.48\columnwidth]{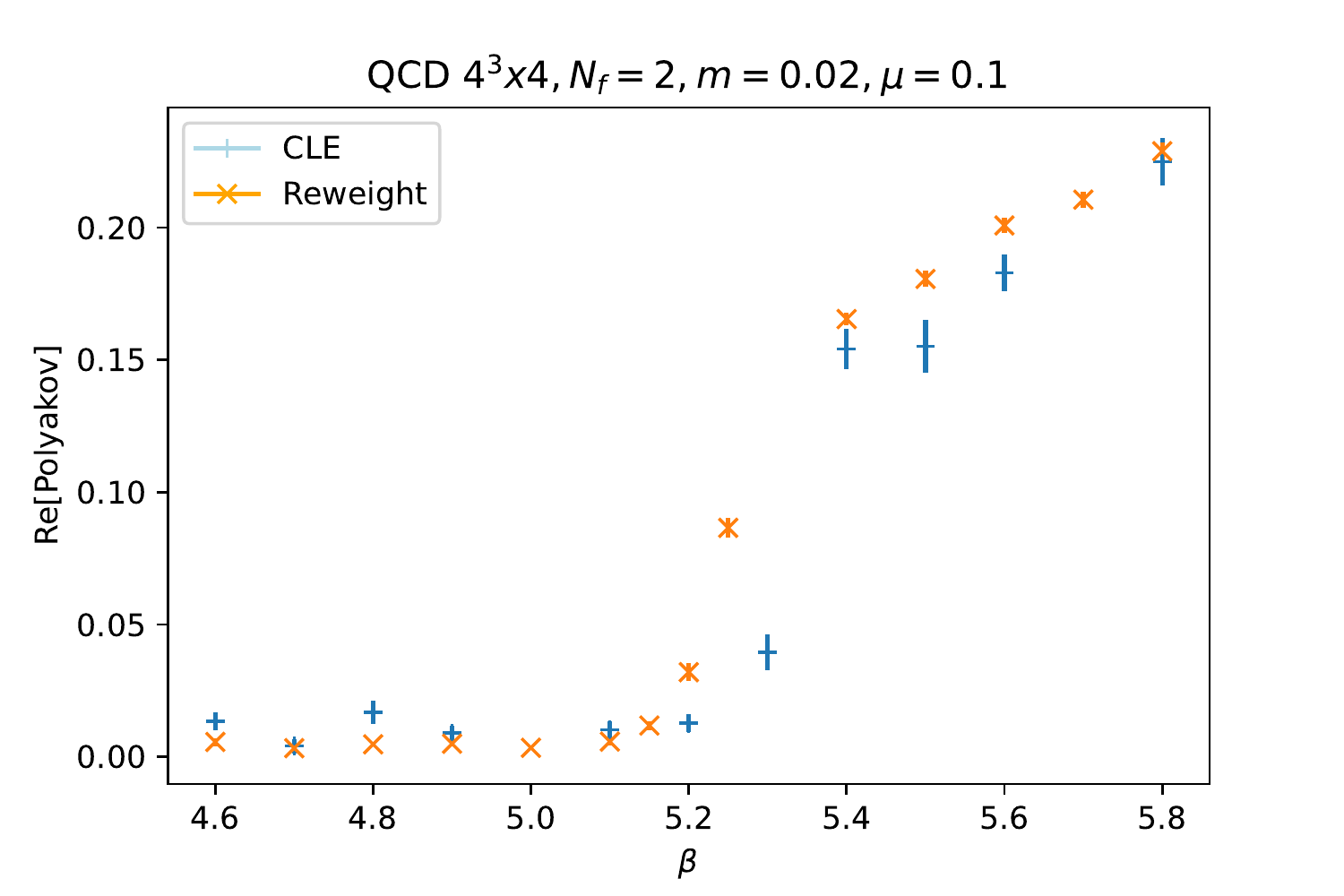}
  \includegraphics[width=0.48\columnwidth]{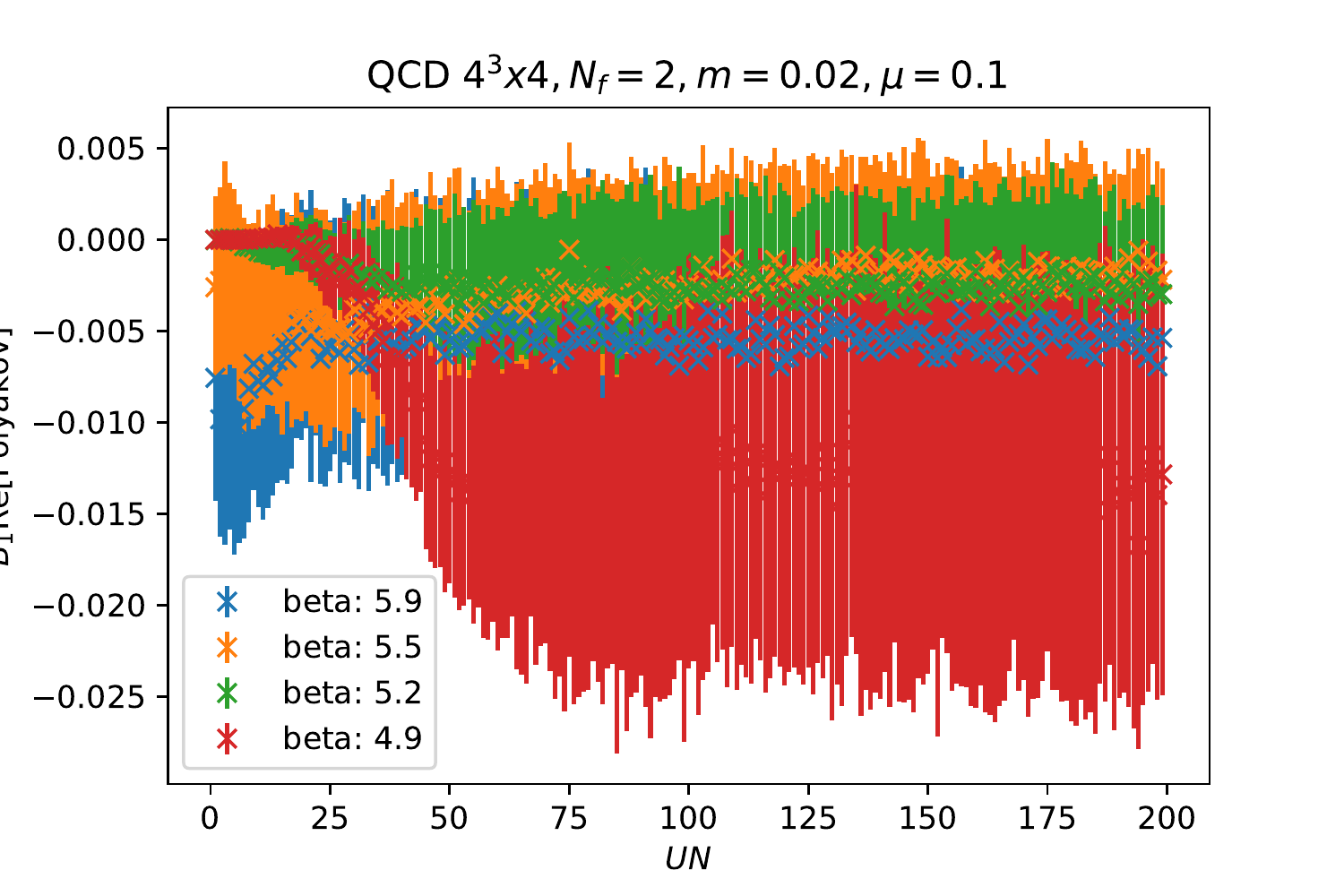}
\caption{On the left panel the Polyakov loop is plotted, with results calculated from CLE(blue) and reweighted HMC(orange). On the right panel the boundary terms are shown for various values of beta as a function of the cutoff. }
\label{pol_mu0.1}
\end{center}
\end{figure}
Similarly, in Fig.~\ref{pol_mu0.1} we see the Polyakov loop comparison and boundary terms.
From these plots it can be seen that the incorrect results are signalled by non-zero boundary terms. In fact the more incorrect the result, the larger  are the magnitude of the boundary terms. 


\section{Dynamical stabilization}
As we can see by the boundary terms, the simulations at low-beta  
yields incorrect results, as they have distributions with slow decay on the complexified manifold.
To negate this problem. a gauge invariant force to the drift term is introduced \cite{Attanasio:2018rtq}:  
\begin{equation}
    K_{x,\mu}^a \rightarrow K_{x,\mu}^a + i \alpha_{DS} M_x^a 
\end{equation}
with the force it self
\begin{equation}
    M_x^a = ib_x^a\left(\sum_c b_x^c b_x^c \right)^3 \qquad \text{ and } \qquad b_x^a = \textrm{Tr} \left[\lambda^a \sum_\nu U_{x,\nu} U_{x,\nu}^\dagger \right]\,.
\end{equation}
Here the $\alpha_{DS}$ is a tunable parameter controlling the strength of the force towards the SU(3) manifold. We also investigated 
a modifaction of the proposal, where the sum over directions has been removed, implying that the $M$-term now also will depend on the direction $\mu$.
\begin{equation}
    {M'}_{x,\mu}^a = ib_x^a\left( 2 \textrm{Tr} \left[ \left( U_{x,\mu} U_{x,\mu}^\dagger \right)^2 \right] - \frac{2}{3} \textrm{Tr} \left[ U_{x,\mu} U_{x,\mu}^\dagger \right]^2 \right)^3\,,
\end{equation}
Here the term has been simplified, using an identity of the Gell-Mann matrices.

\section{Results with dynamical stabilization}
In this section we compare the results of dynamically stabilized CLE simulations to results of HMC simulations at zero quark chemical potential$(\mu =0)$  reweighted to a non-zero $(\mu = 0.1)$. The CLE simulations tend to crash due to infinities, when using low $\beta$ parameters (low temperatures). Therefore we show results for nonzero $\alpha$ parameters as dynamical stabilization helps suppressing this instable behavior.  
We used $M$ and $M'$ as defined in the previous section, and we have observed that mostly it doesn't really matter which one we use. So from now on we present the data that has been found using the $M$-term, as described in \cite{Attanasio:2018rtq}.  In this section, we present results on $8^3 \times 4$ lattices, and here the noisy estimator \cite{Sexty:2013ica} has been used to calculate the fermionic drift terms.

\begin{figure}[!htb]
\begin{center}
  \includegraphics[width=0.41\columnwidth]{new_plots/sp_mu0.1.pdf}
  \includegraphics[width=0.41\columnwidth]{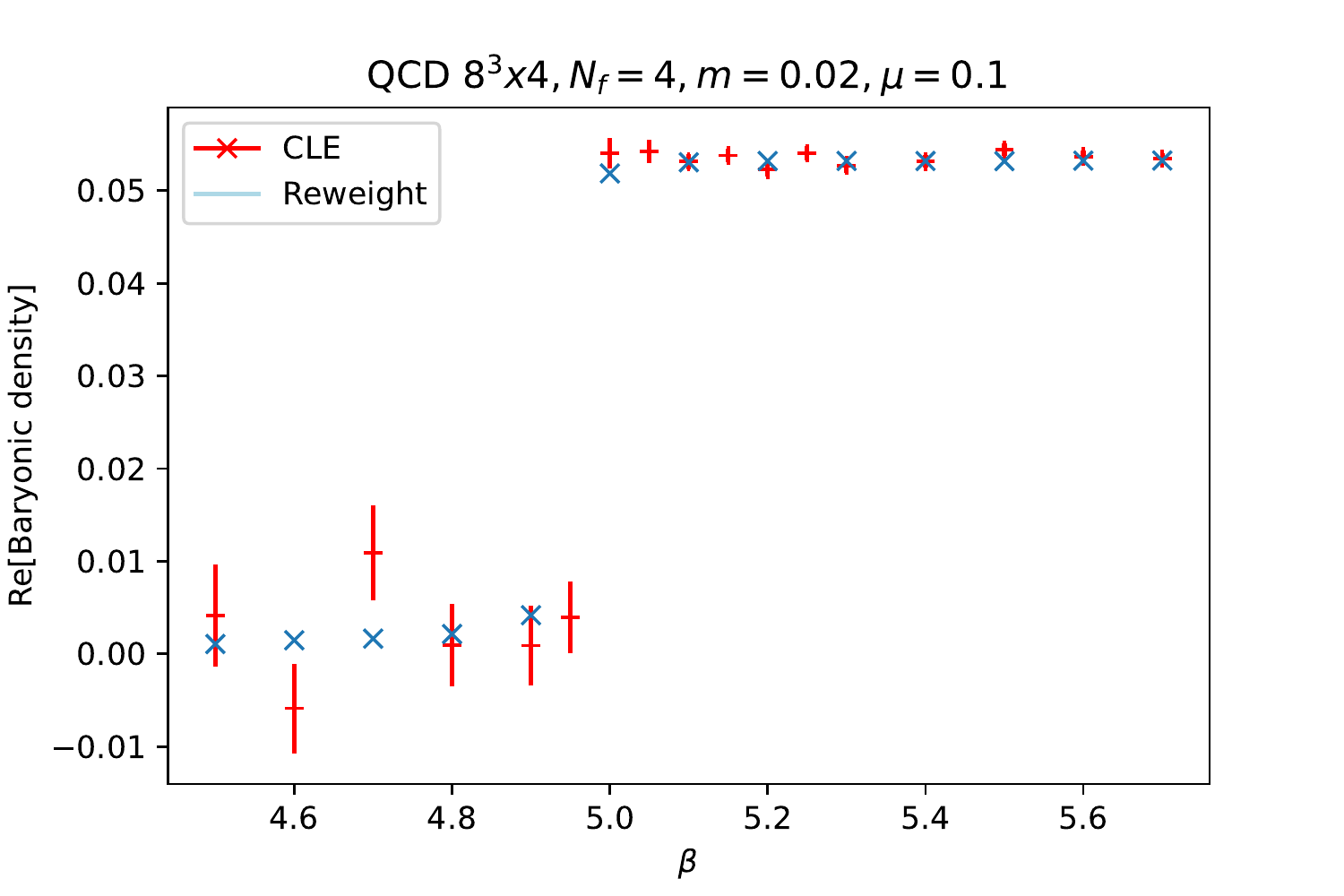}
  \includegraphics[width=0.41\columnwidth]{new_plots/pol_mu0.1.pdf}
  \includegraphics[width=0.41\columnwidth]{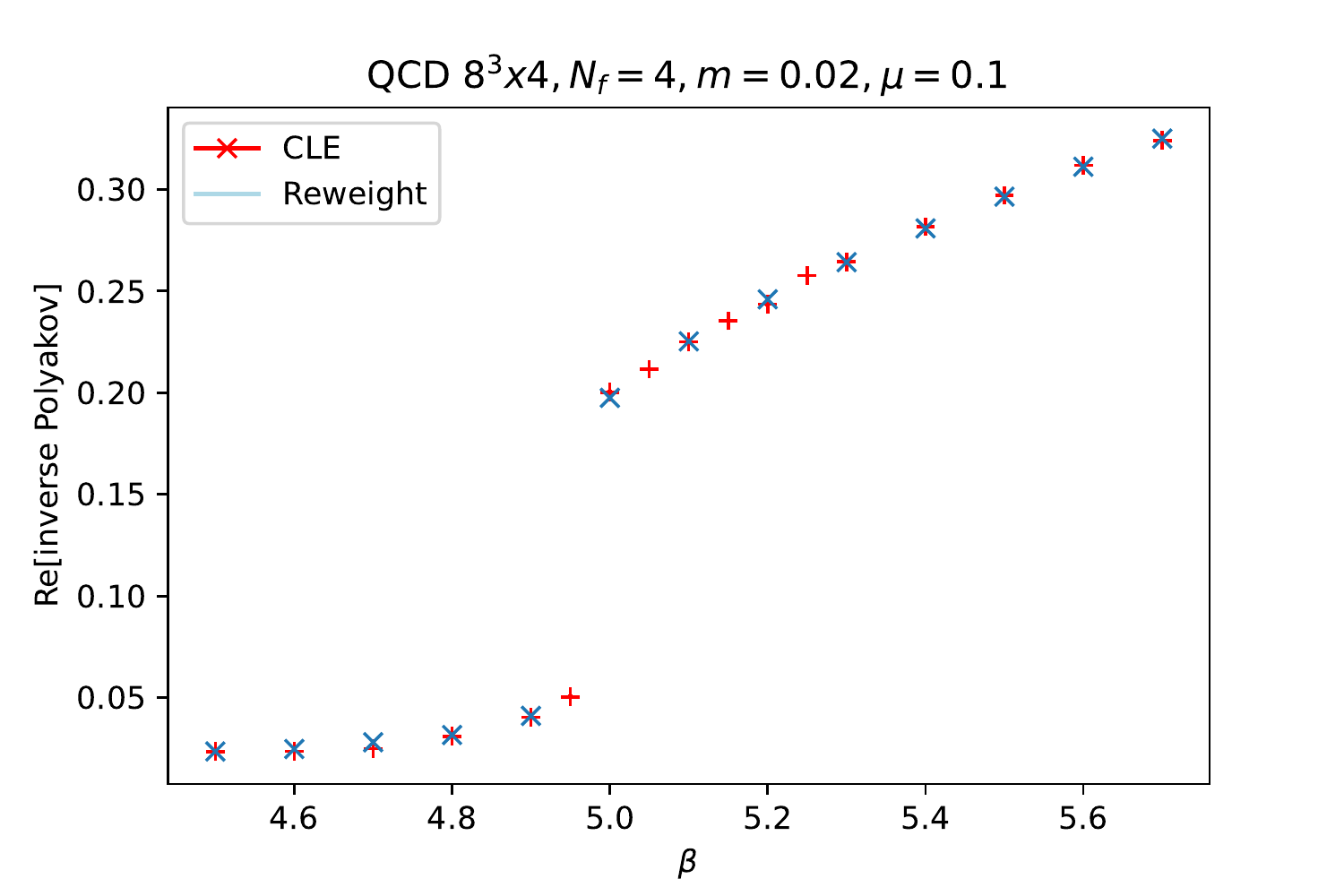}
\caption{Comparison between CLE, and reweighted HMC data, for the four observables studied in this paper. Top left: Spatial Plaquette. Top right: Baryonic density. Lower left: Polyakov loop. Lower right: Inverse Polyakov loop. The CLE data is simulated with $\alpha_{DS} = 10^6$.}
\label{rews}
\end{center}
\end{figure}

\begin{figure}[!htb]
\begin{center}
    \includegraphics[width=0.42\columnwidth]{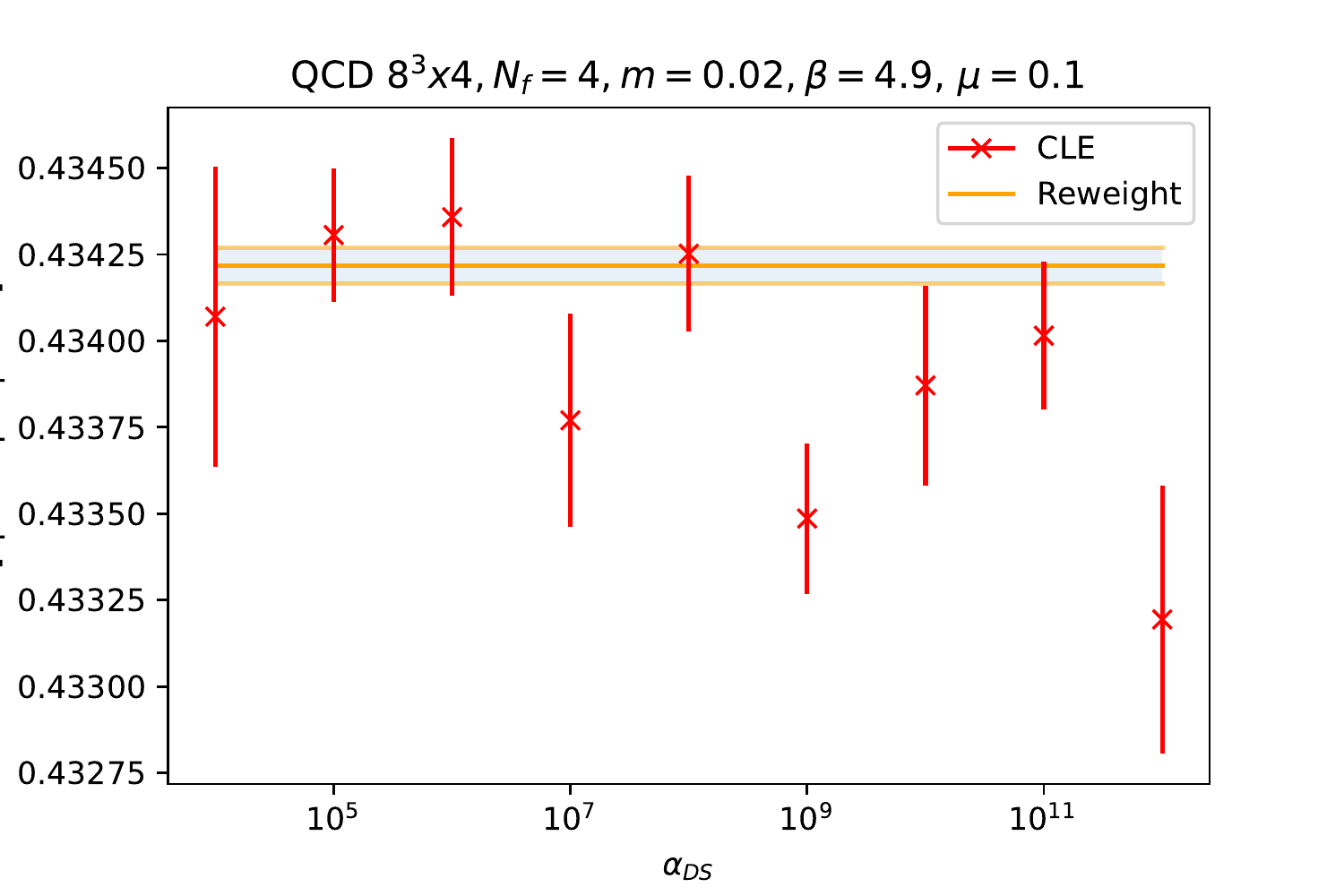}
    \includegraphics[width=0.42\columnwidth]{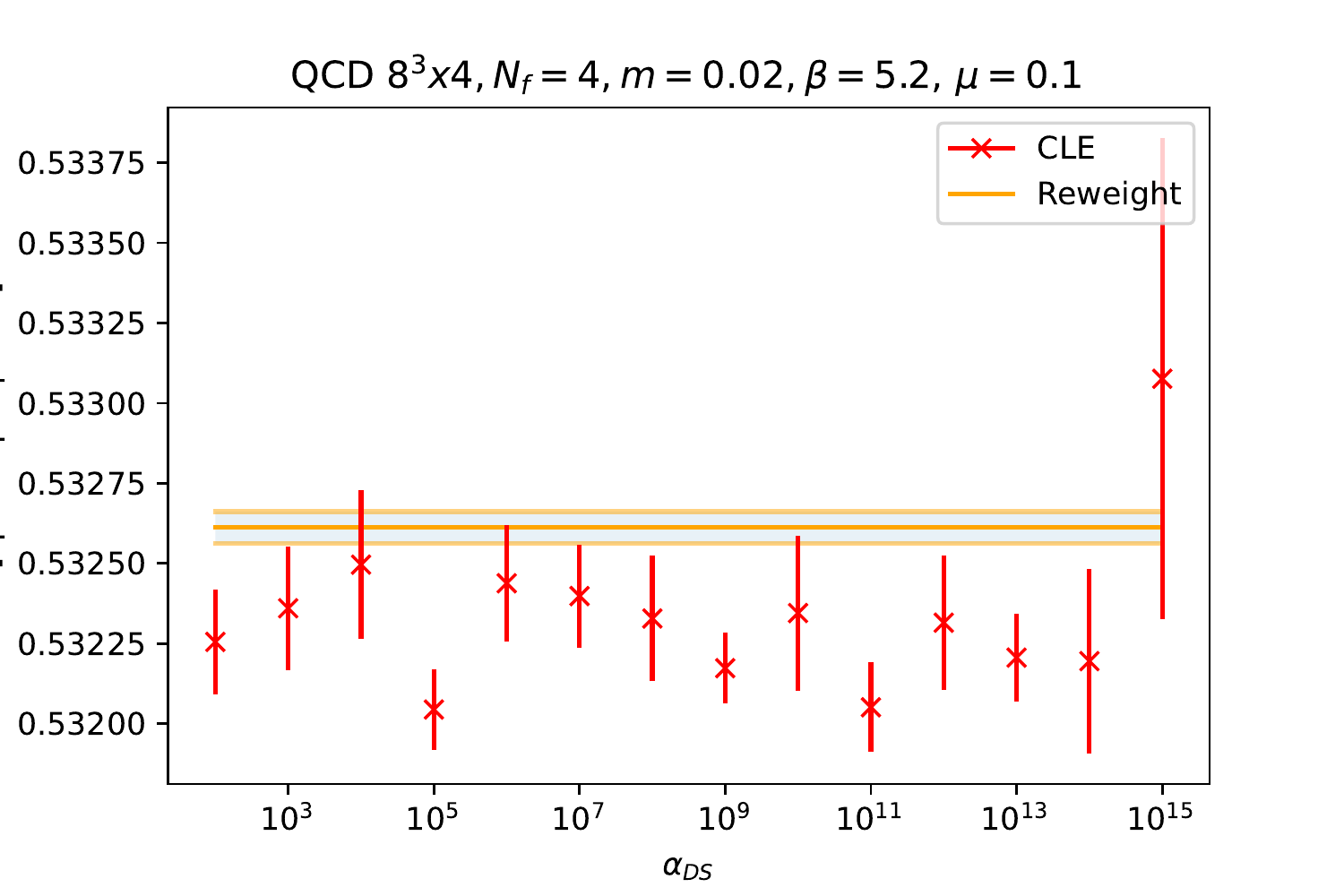}   
    \includegraphics[width=0.42\columnwidth]{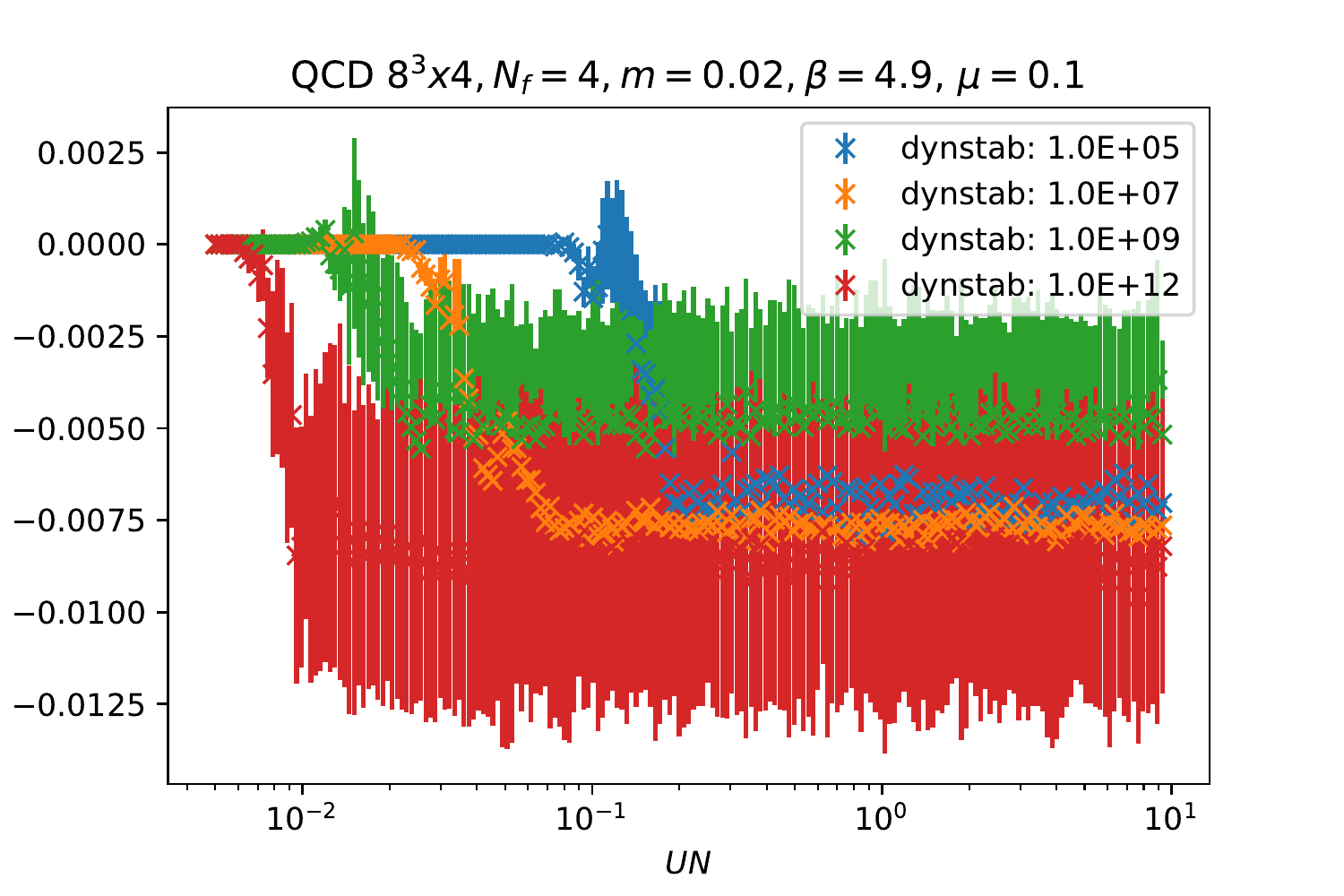}
    \includegraphics[width=0.42\columnwidth]{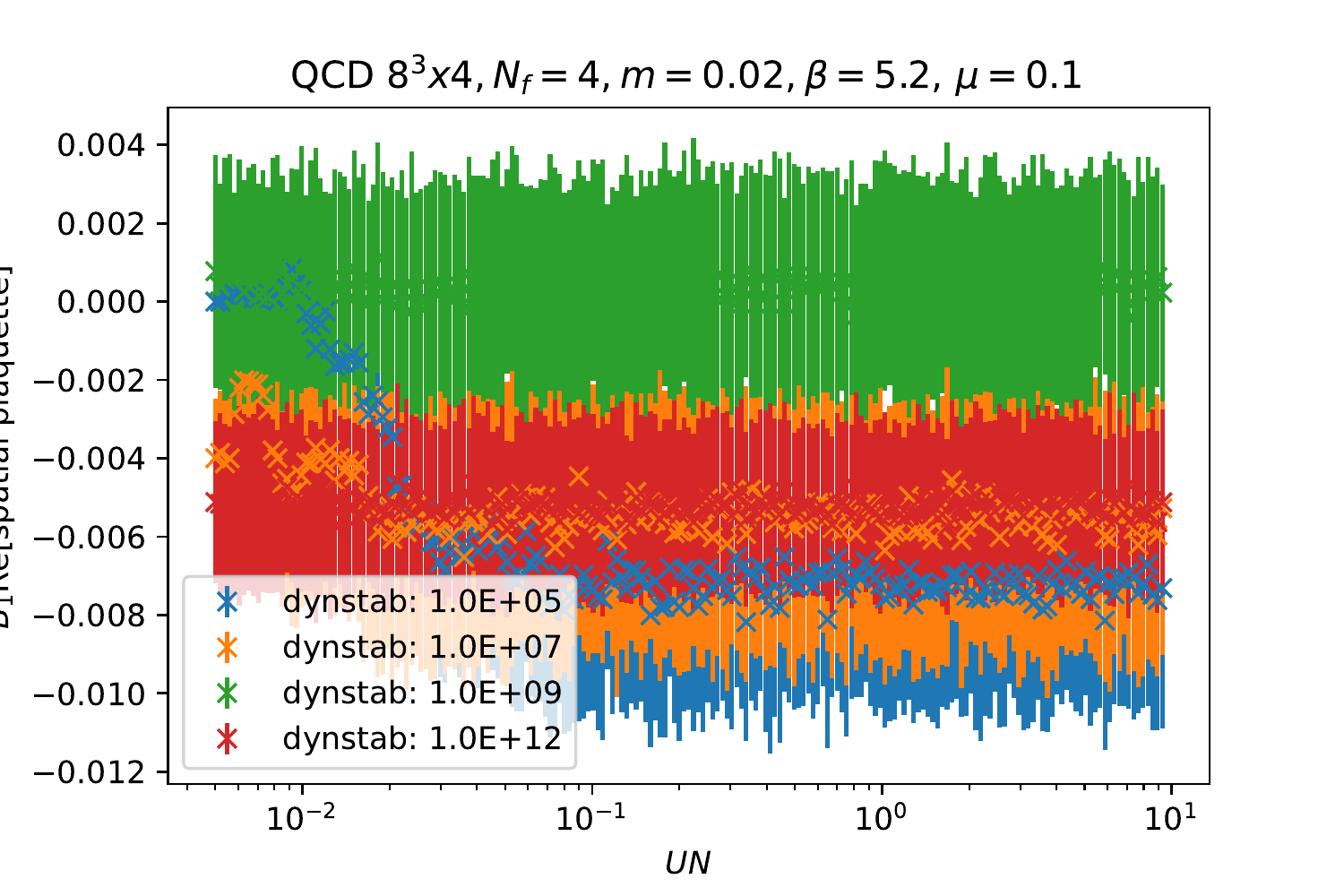}
\caption{The spatial plaquette(top) and its boundary term(bottom) for low(left) temperature and high(right) temperature.}
\label{sp_84}
\end{center}
\end{figure}

\begin{figure}[!htb]
\begin{center}
    \includegraphics[width=0.45\columnwidth]{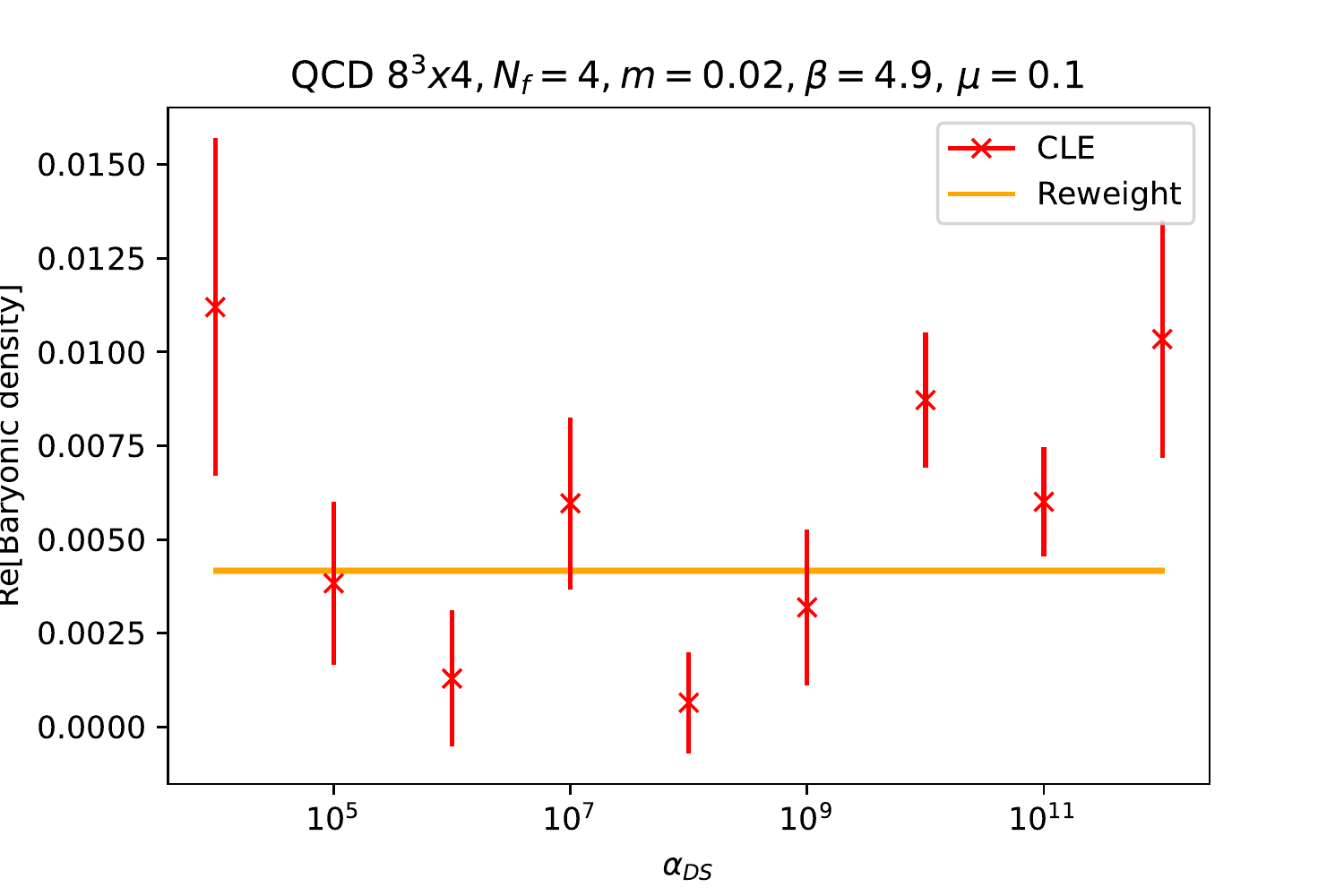}
    \includegraphics[width=0.45\columnwidth]{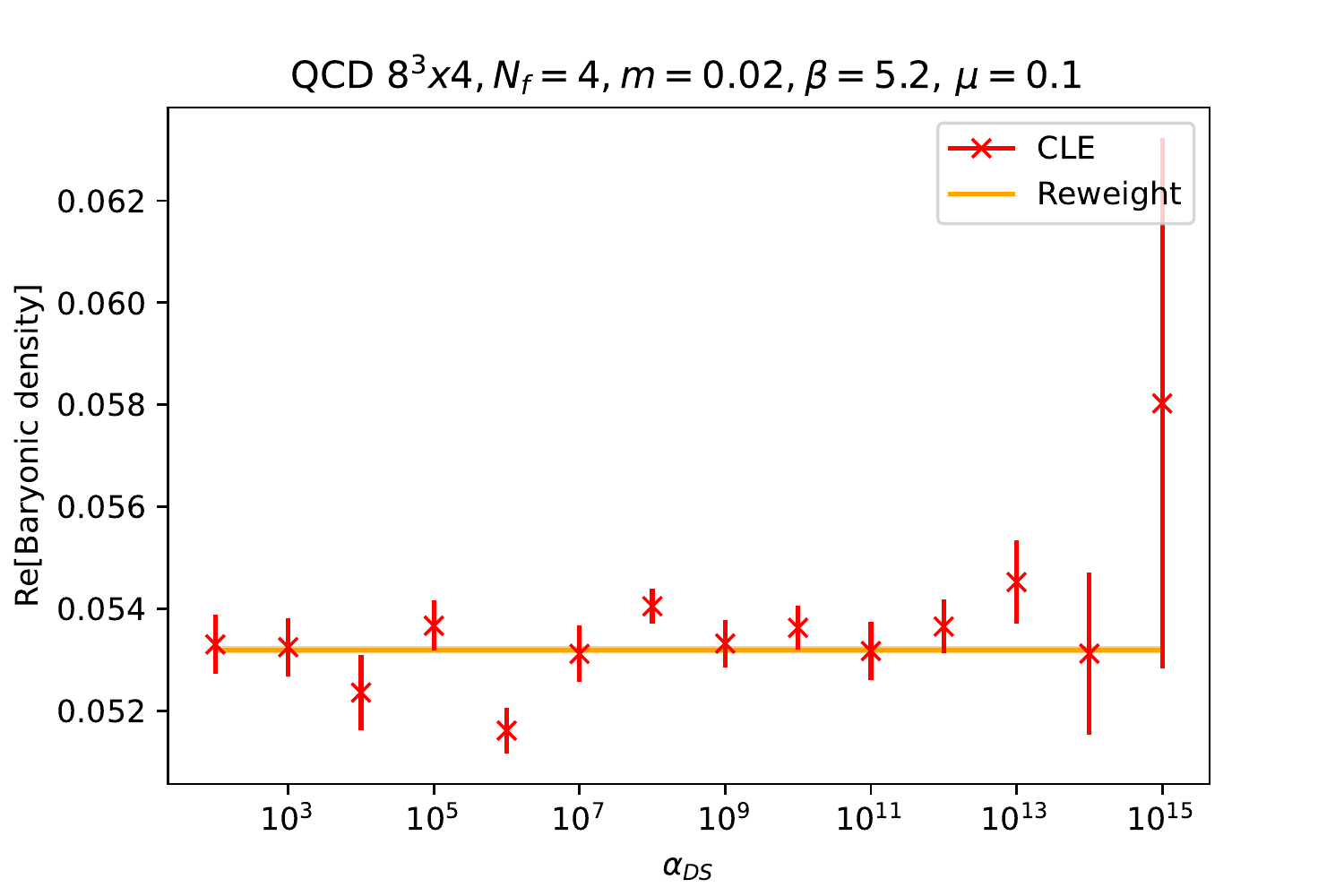}
    \includegraphics[width=0.45\columnwidth]{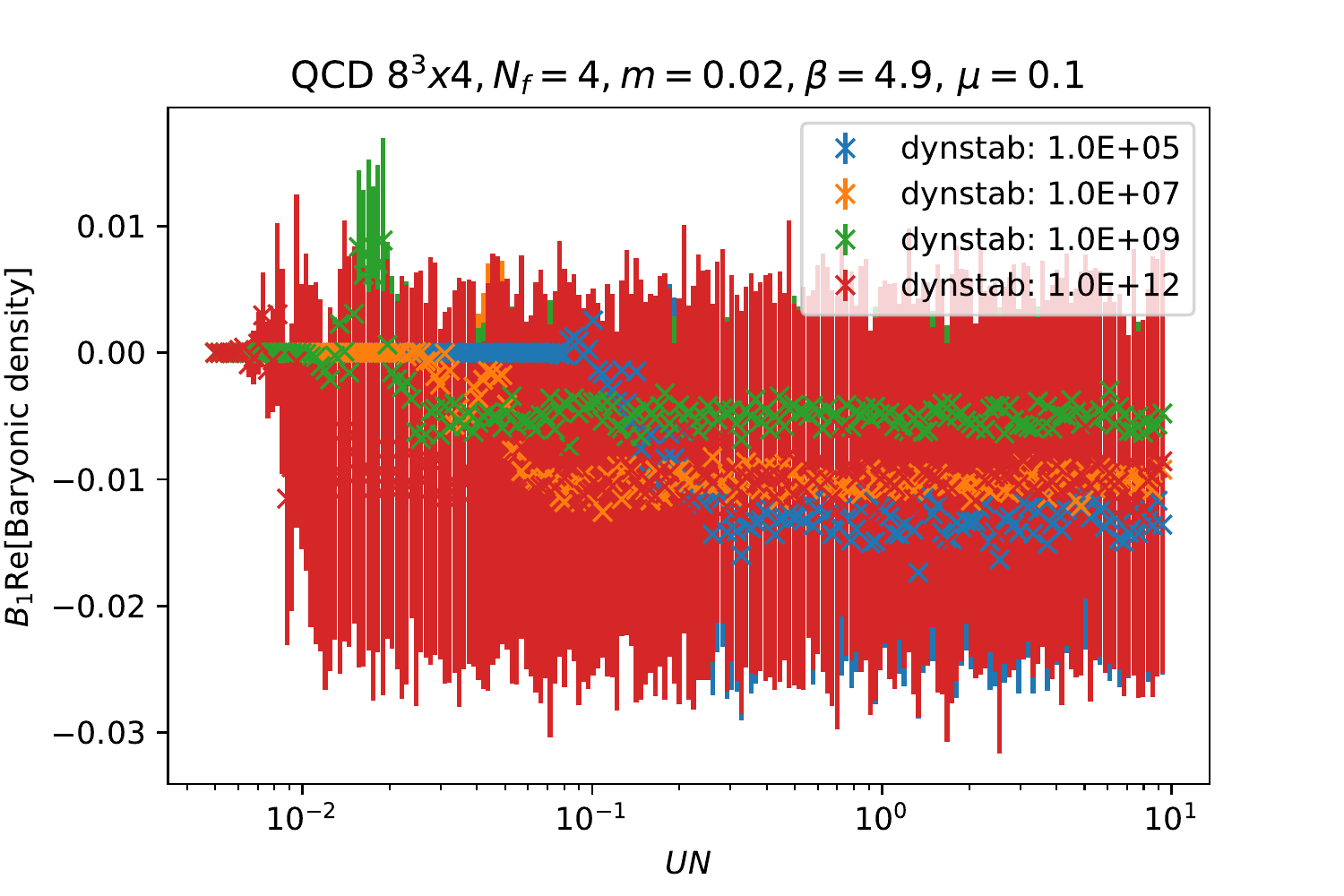}
    \includegraphics[width=0.45\columnwidth]{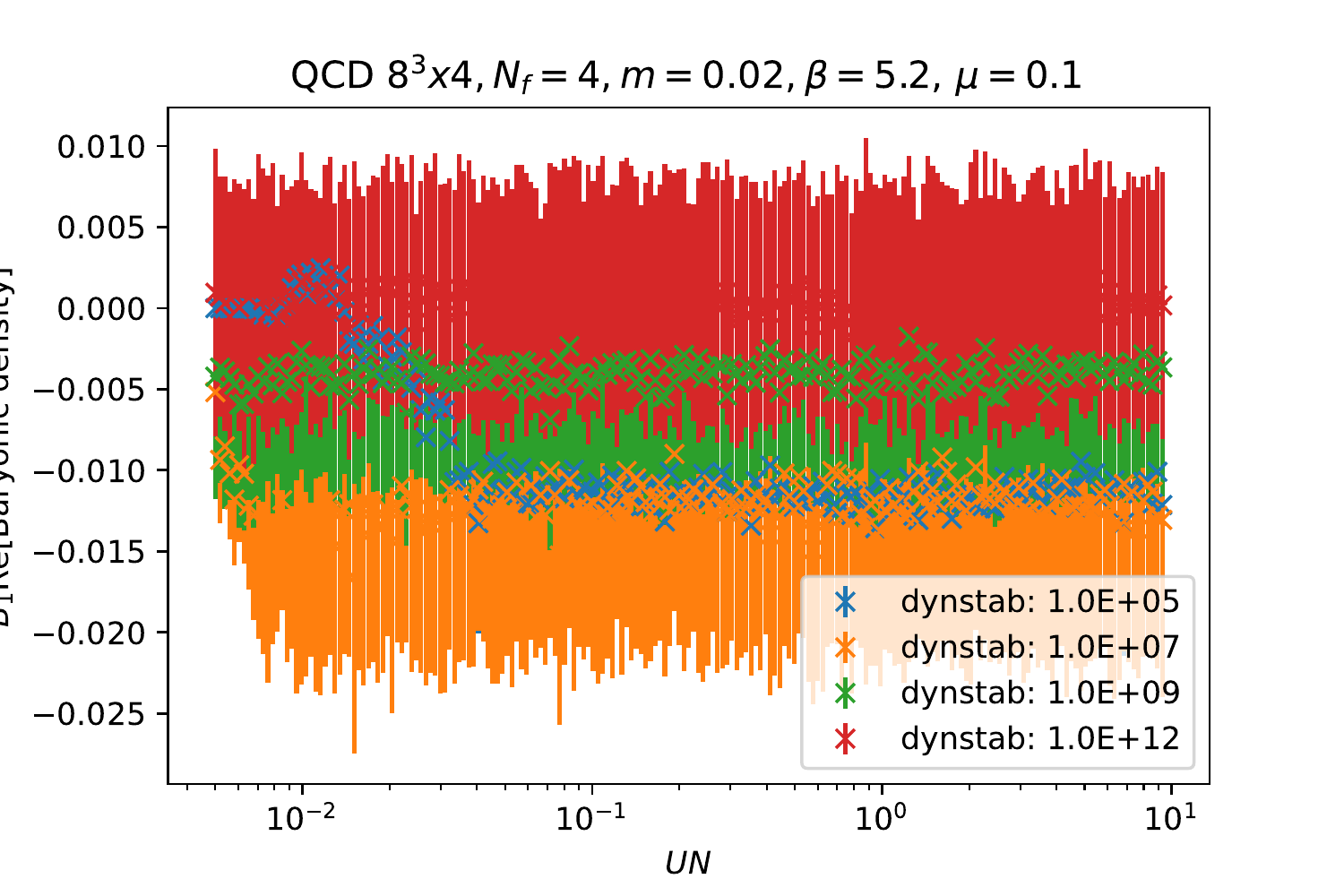}
\caption{The baryonic density(top) and its boundary term(bottom) for low(left) temperature and high(right) temperature. }
\label{pol_84}
\end{center}
\end{figure}

\begin{figure}[!htb]
\begin{center}
  \includegraphics[width=0.45\columnwidth]{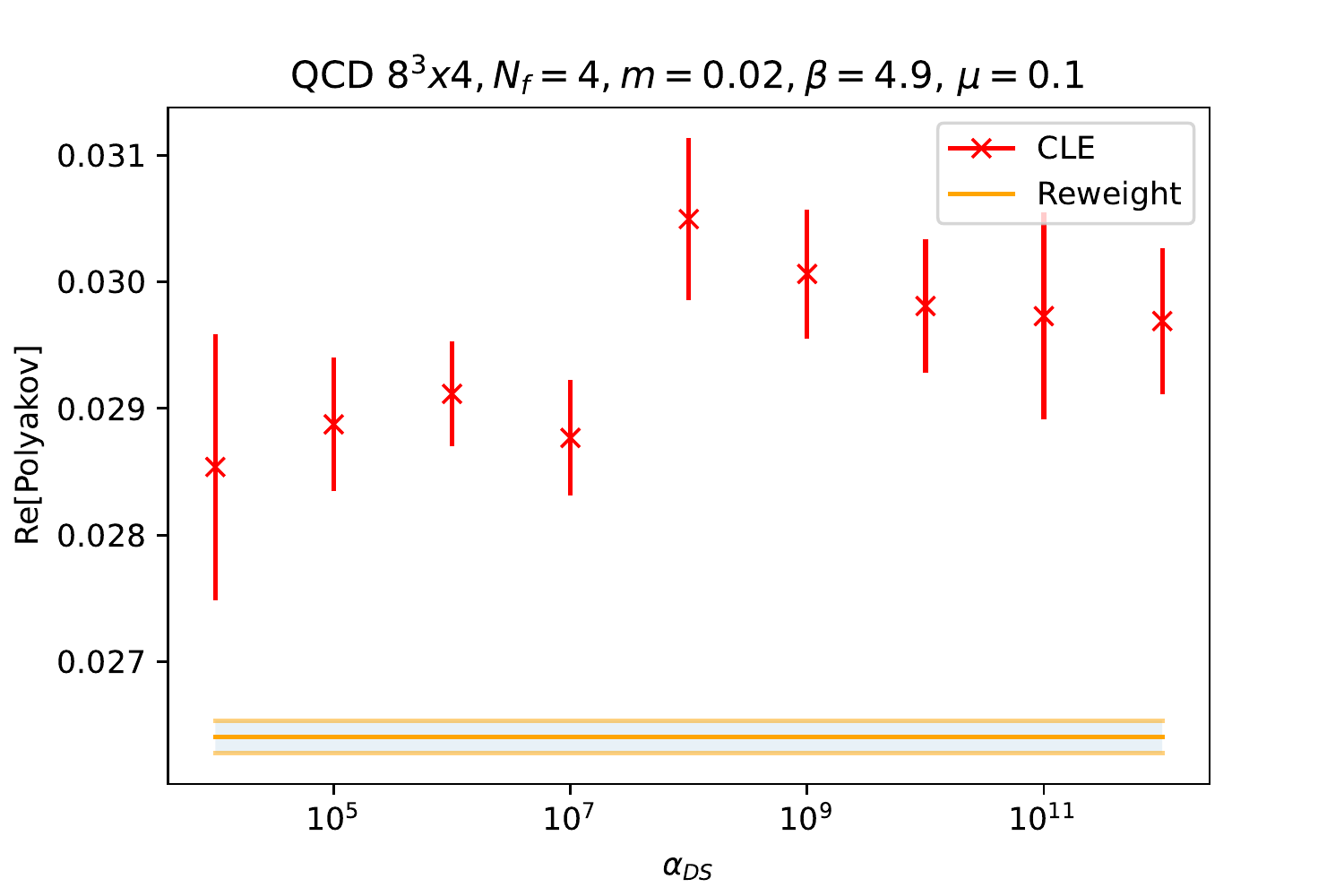}
  \includegraphics[width=0.45\columnwidth]{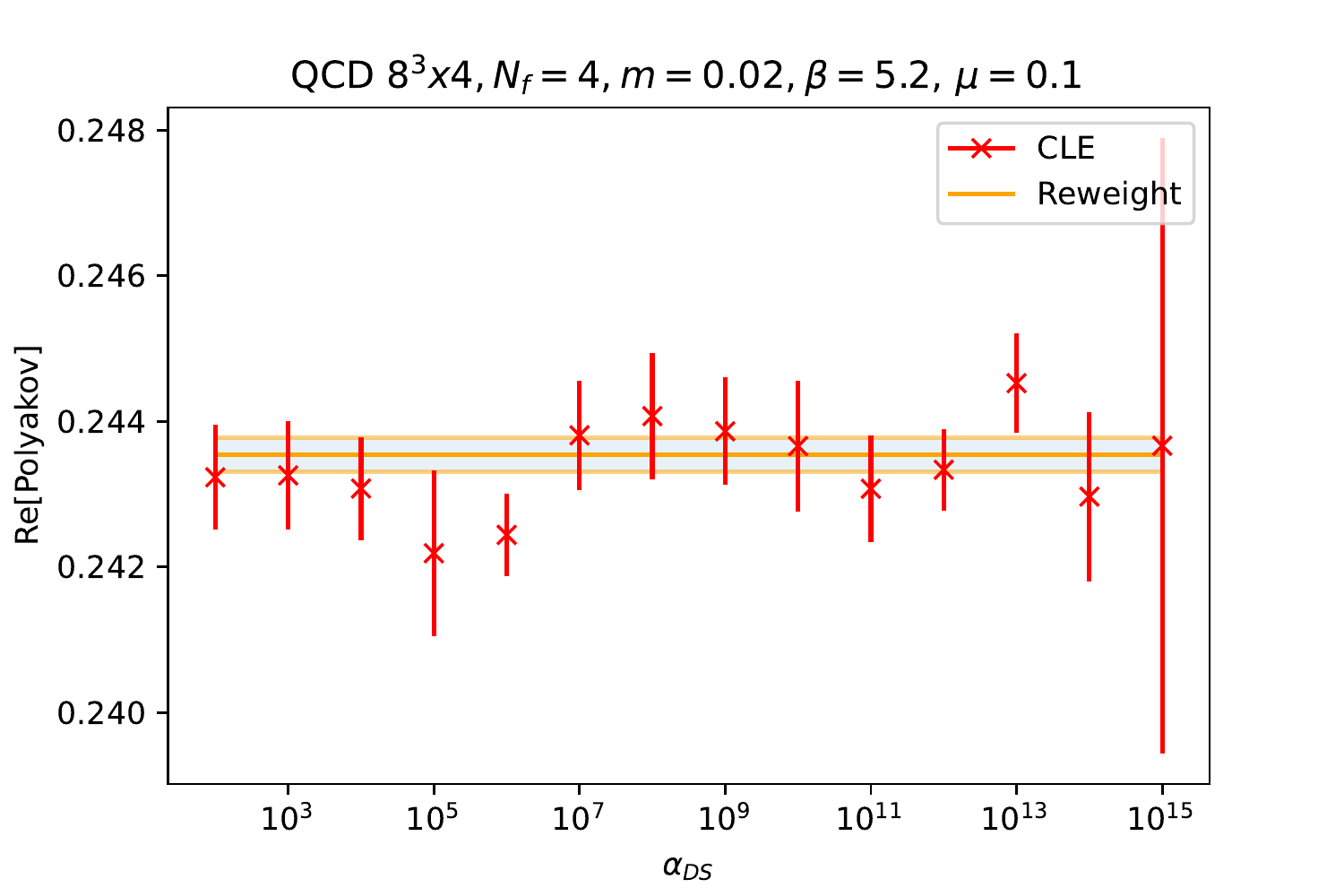}
  \includegraphics[width=0.45\columnwidth]{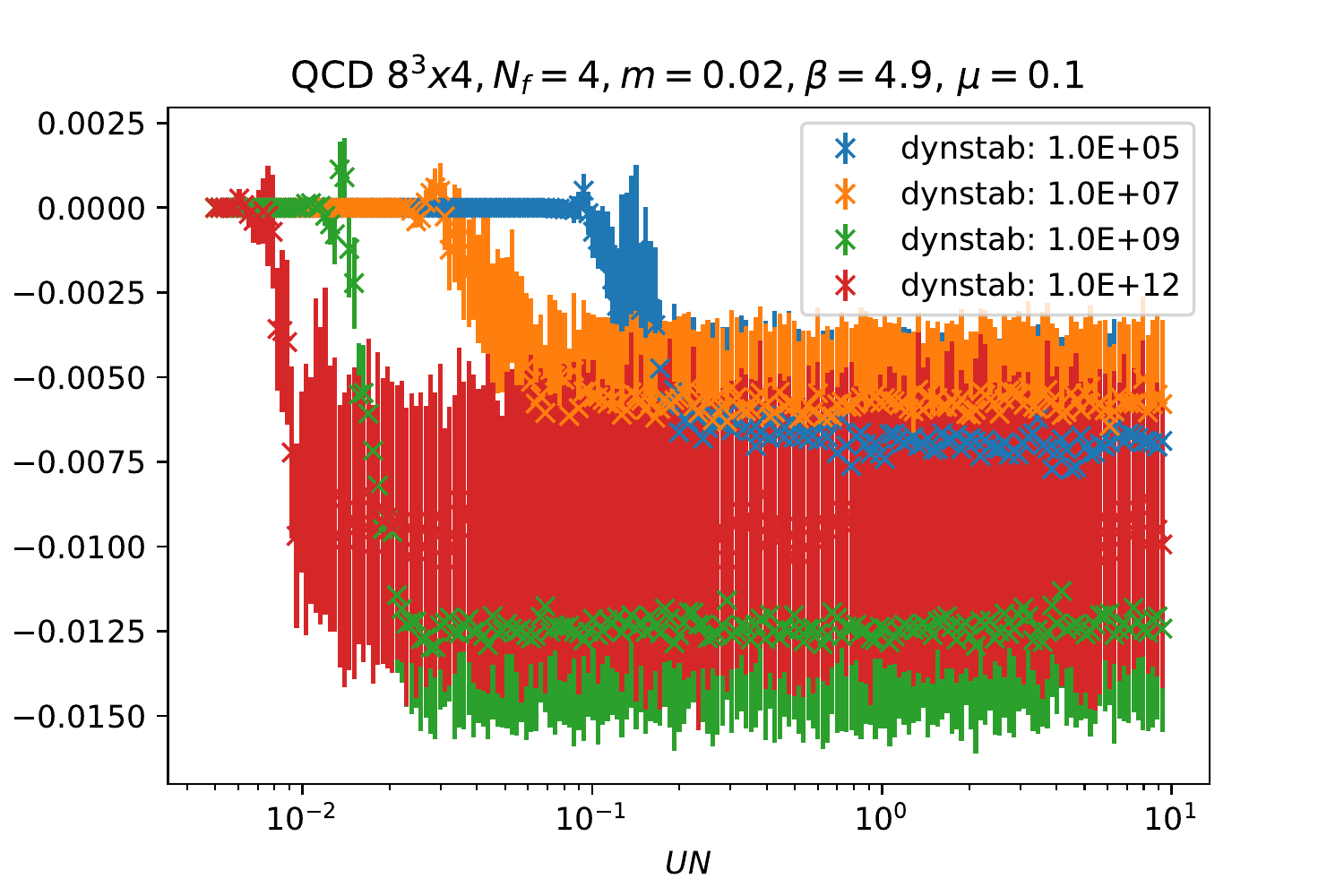}
  \includegraphics[width=0.45\columnwidth]{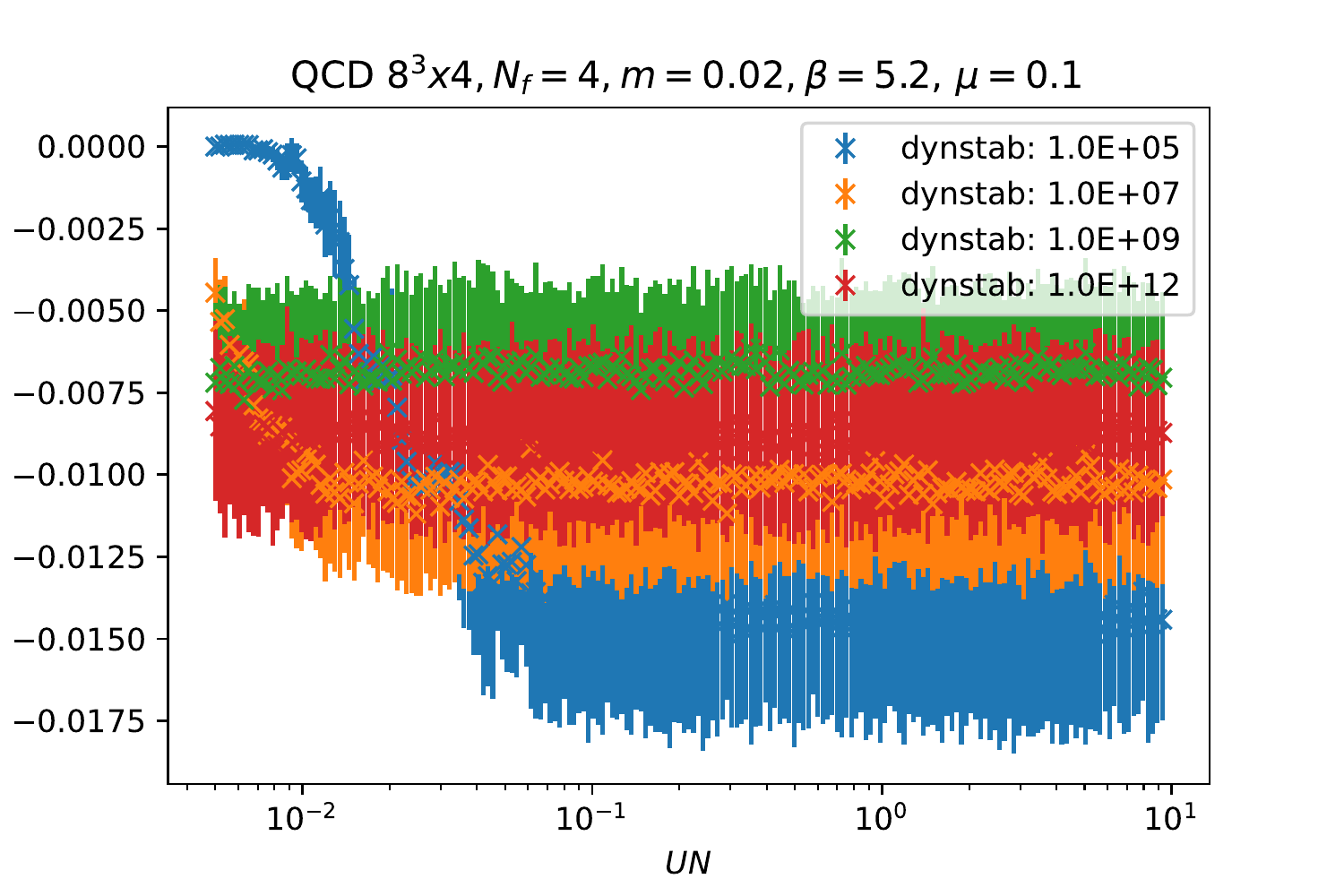}
\caption{The Polyakov loop(top) and its boundary term(bottom) for low(left) temperature and high(right) temperature.}
\label{dens_84}
\end{center}
\end{figure}

\begin{figure}[!htb]
\begin{center}
  \includegraphics[width=0.45\columnwidth]{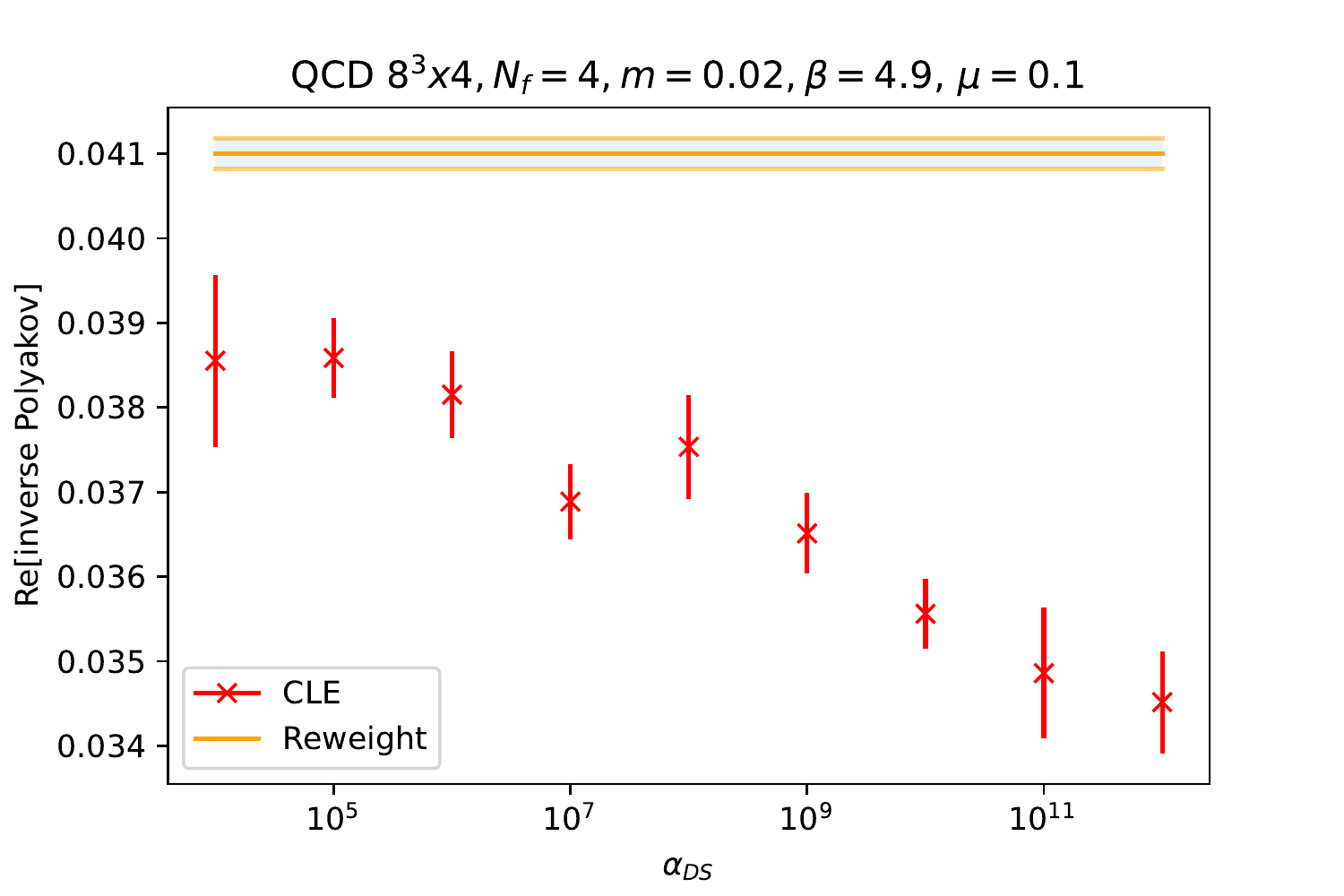}
  \includegraphics[width=0.45\columnwidth]{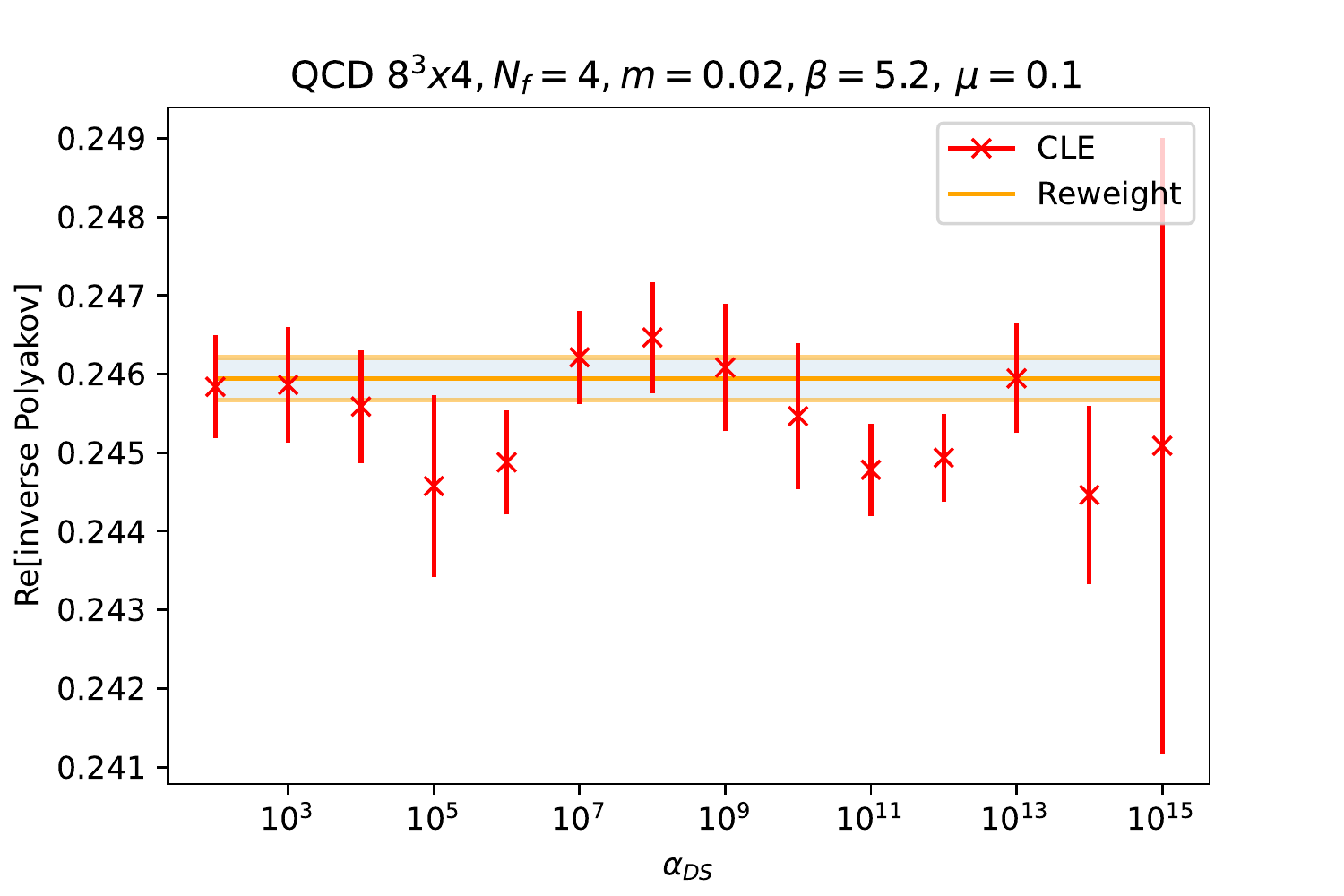}
  \includegraphics[width=0.45\columnwidth]{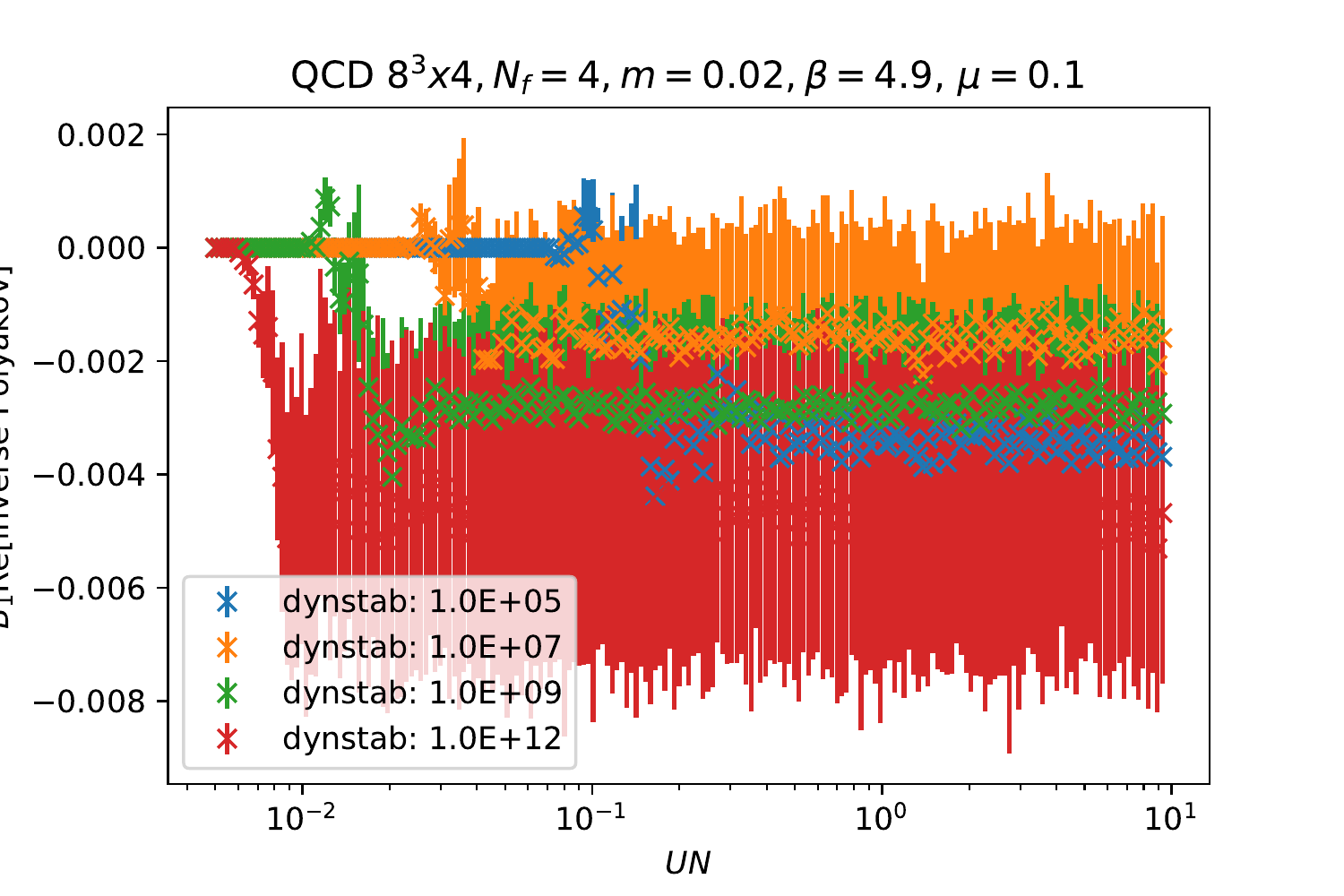}
  \includegraphics[width=0.45\columnwidth]{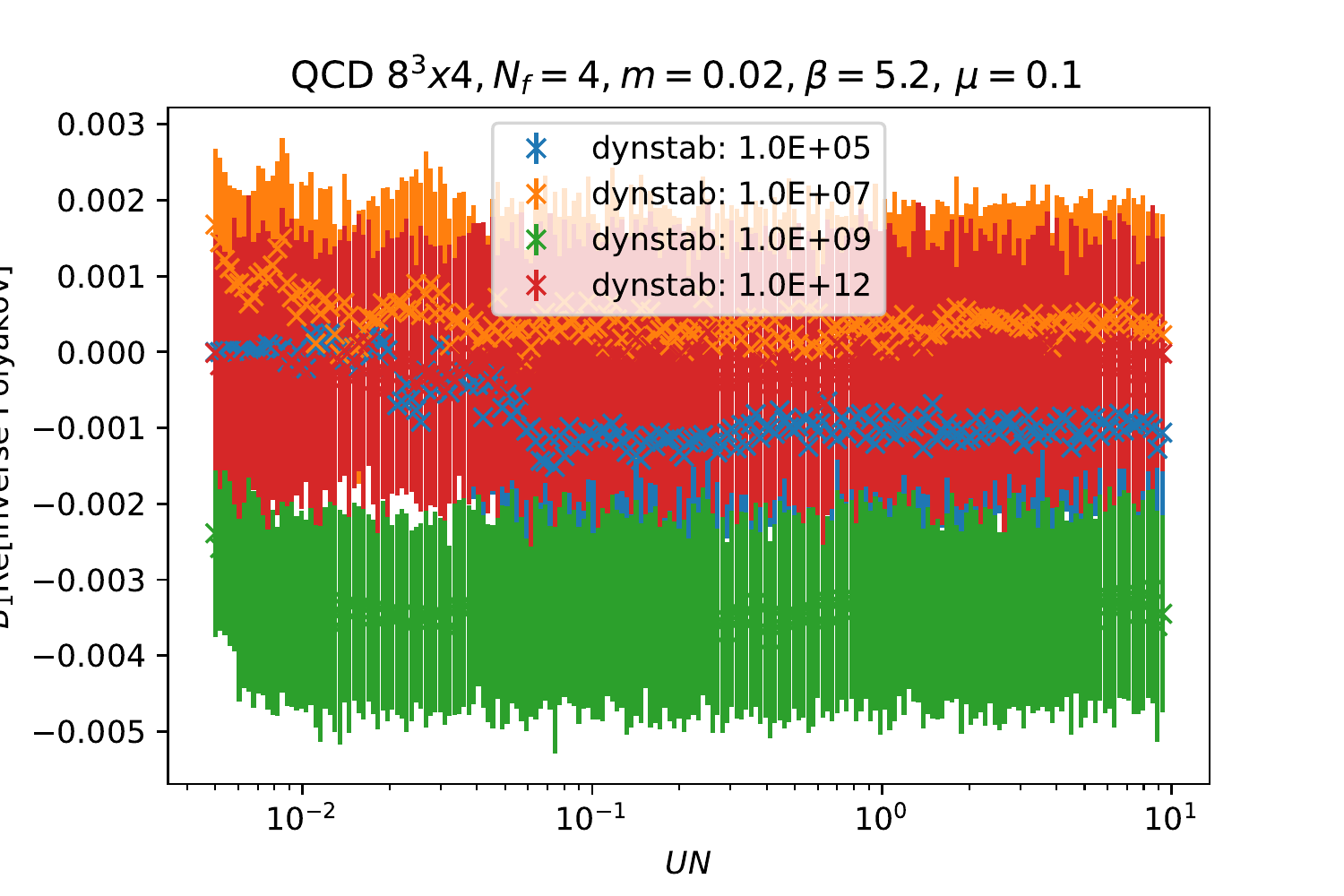}
\caption{The inverse Polyakov loop(top) and its boundary term(bottom) for low(left) temperature and high(right) temperature.}
\label{polinv_84}
\end{center}
\end{figure}

In \Cref{rews} we show the spatial plaquette average, Polyakov loops and its inverse and the baryonic density, calculated with CLE simulations using dynamical stabilization with $ \alpha=10^6$ and reweighted HMC results. One observes good overall agreement even at low $\beta$ parameters.
In ~\Cref{sp_84,pol_84,dens_84,polinv_84} we show the results of the CLE simulations as a function of the $\alpha$ parameter, where the reweighted HMC result is also indicated, as well as the boundary terms of the observables.
We use two $\beta$ parameters, one in the low temperature and one in the high temperature phase.  We observe that at high temperatures the results are correct. In this case actually simulations without dynamical 
stabilization behave similarly well. At low temperatures, we observe that the spatial plaquette and density observables within errors
reproduce the reweighted results, whereas we see discrepancies in the Polyakov loop and its inverse. Although the magnitude of these errors 
are quite small they seem to be statistically significant. The boundary terms are quite small for all observables and temperatures (much smaller in magnitude as for the $\alpha=0$ simulations at low $\beta$).
Further investigation is neccessary to be able to verify whether the small discrepancy in the Polyakov loops is accurately signalled by 
the boundary terms. Note that the new term introduced in the Langevin equation is non-holomorphic, so in principle 
the boundary term analysis is invalidated in this setup, although in practice it seems to give results consistent with expectations. 


\newpage

\section{Conclusions}
We have studied the boundary terms in full QCD simulations using the Complex Langevin equation.
We observe a good performance of CLE for high temperatures, where dynamical stabilization is not needed, and we obtain correct results, as confirmed by comparison to reweighting and correctly signalled by vanishing boundary terms. At low inverse coupling however, incorrect results
are obtained, which are correctly signalled by non-vanishing boundary terms. We have tested dynamical stabilisation by comparing its results 
to reweighting while also measuring boundary terms of some observables. We observe correct results at high temperatures, and mostly correct results at low temperatures (the Polyakov loop shows a small discrepancy). This study will be presented in more detail in an upcoming article \cite{newpaper}.

\bibliographystyle{apsrev}
\bibliography{cites} 

\end{document}